\documentclass[final,twocolumn]{article}

\usepackage[T1]{fontenc}
\usepackage{array}
\usepackage{lmodern}
\usepackage{caption}
\usepackage{subcaption}
\usepackage{authblk}
\usepackage{amsmath}
\usepackage{amssymb}
\usepackage[pdftex]{graphicx}
\usepackage{multirow}
\usepackage{color}
\usepackage{hyperref}
\usepackage{graphicx}
\usepackage[latin1]{inputenc}

\author[1,2]{P. Agostinetti}
\author[1]{S. Dal Bello}
\author[3]{F. Dinh}
\author[3]{A. Ferrara}
\author[1]{M. Fincato}
\author[1,2]{L. Grando}
\author[3]{M. Mura}
\author[1,2]{A. Murari}
\author[1,4]{E. Sartori}
\author[1]{M. Siragusa}
\author[3]{F. Siviero}
\author[1,5]{F. Veronese}

\affil[1]{Consorzio RFX (CNR, ENEA, INFN, Universit\`a di Padova, Acciaierie Venete SpA), Corso Stati Uniti 4, 35127 Padova, Italy}
\affil[2]{Institute for Plasma Science and Tecnology-Section of Padova, Corso Stati Uniti 4, 35127 Padova, Italy}
\affil[3]{SAES Getters SpA, Viale Italia 77, 20045 Lainate (MI), Italy}
\affil[4]{Dept. of Management and Engineering, University of Padova, Stradella S. Nicola, 3, 36100 Vicenza, Italy}
\affil[5]{Dept. of Electrical Engineering, University of Padova, Via Gradenigo 6/A, 35131 Padova, Italy}
\date{\url{https://doi.org/10.1016/j.fusengdes.2023.113638}}

\title{Conceptual design of the Gas Injection and Vacuum System for DTT NBI}
\begin{document}
\twocolumn[
	\begin{@twocolumnfalse}
		\maketitle
		\begin{abstract}
		The Divertor Tokamak Test (DTT) is a new experimental facility whose construction is starting in Frascati, Rome, Italy; its main goals are improving the understanding of plasma-wall interactions and supporting the development of ITER and DEMO. DTT will be equipped with a Neutral Beam Injector (NBI) based on negative deuterium ions, designed to inject 10 MW of power to the tokamak.

A fundamental system for the good operations of the DTT NBI will be its Gas injection and Vacuum System (GVS). Indeed, the efficiency of the entire NBI strongly depends on the good performance of its GVS.

The GVS for DTT NBI will be composed of two systems working in parallel: a grounded section connected to the main vacuum vessel, and a high voltage part connected to the ion source vessel and working at -510 kV voltage. The grounded part will feature a fore vacuum system (given by screw and roots pumps) plus a high vacuum system based on turbo-molecular pumps located on the side walls of the vessel and Non-Evaporable Getter (NEG) pumps located inside the vessel on the upper and lower surfaces. On the other hand, the high voltage part will feature a fore vacuum system (given by two compact screw pumps mounted on the external surface for the ion source vessel) plus a high vacuum system based on turbo-molecular pumps also located on the sidewalls of the ion source vessel. A dedicated deuterium gas injection will feed the process gas to the ion source and the neutralizer.

This paper gives a description of the conceptual design of the GVS for DTT NBI, and of the procedure followed to optimize this system considering the operational requirements and the other constraints of the DTT NBI.
\\
		\end{abstract}
	\end{@twocolumnfalse}
	]
%

\let\thefootnote\relax\footnotetext{E-mail: \url{piero.agostinetti@igi.cnr.it}}

\section*{Introduction}
\label{Introduction}

        \begin{figure*}
        \centerline{\includegraphics[width=0.75 \textwidth]{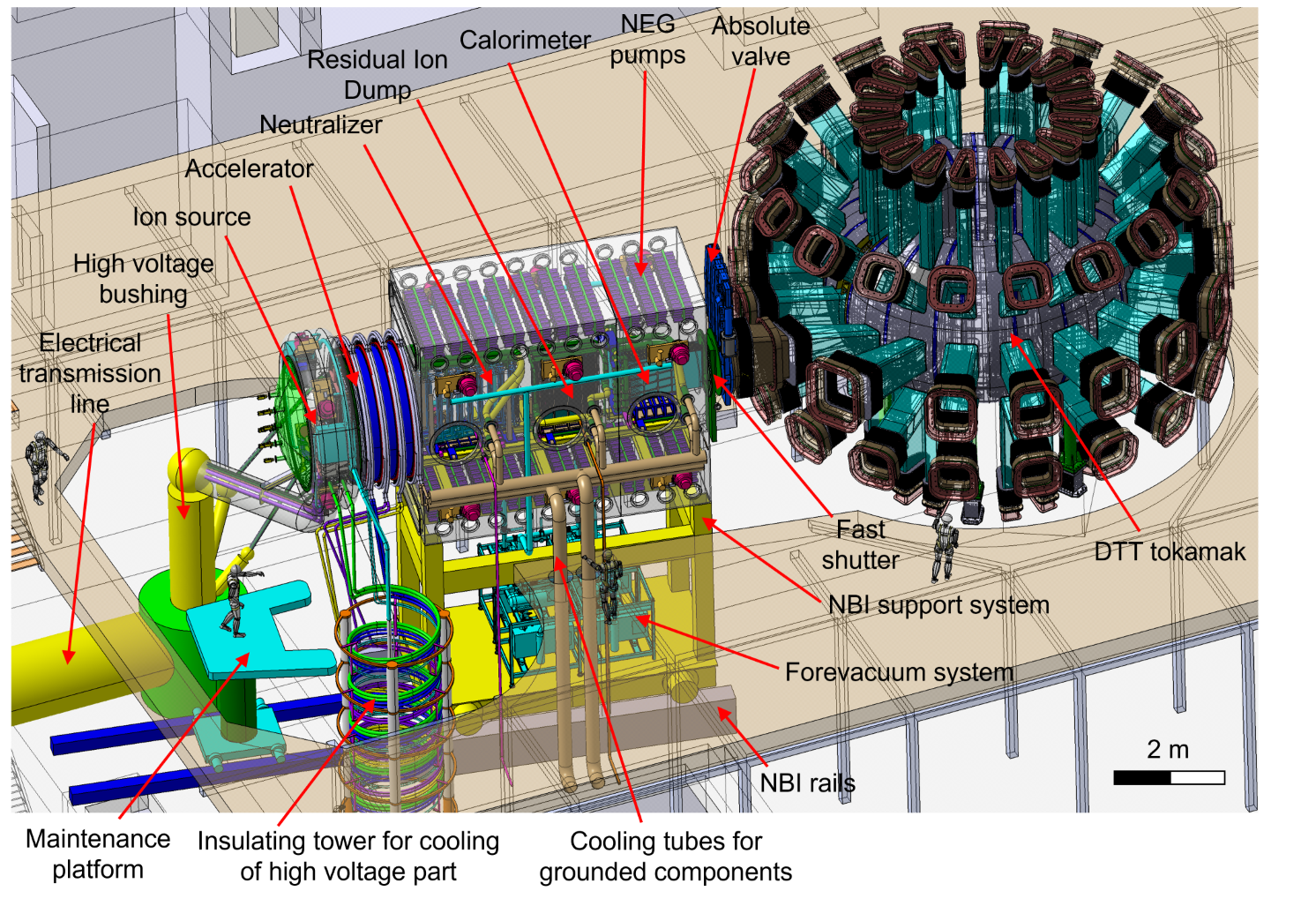}}
        \caption{Overview of the conceptual design of the beamline for the DTT NBI (2022 status).}
        \label{Fig_general}
        \end{figure*}

The main purpose of the Divertor Tokamak Test facility (DTT) is to study solutions to mitigate the issue of power exhaust in reactor relevant conditions \cite{cit1}, particularly exploring alternative power exhaust solutions for the machines that will follow ITER \cite{cit2,cit3}, in particular the demonstrative power plant DEMO that is currently in the conceptual design phase \cite{cit4}.
In this context, the principal objective of DTT is to mitigate the risk of a difficult extrapolation to a fusion reactor of the conventional divertor based on detached conditions, which will be tested in ITER. The task implies the study of the completely integrated power exhaust problems and the demonstration of how the possible implemented solutions (e.g., advanced divertor configurations or liquid metals) could be integrated in the DEMO device and other fusion reactors.
The key feature of such a study is to equip the machine with a significant amount of auxiliary heating power in order to test innovative divertor concepts. DTT will be able to explore various magnetic divertor configurations and in order to reach a reactor relevant power flow to the divertor, 45 MW of auxiliary power will have to be coupled to the plasma using the following systems: Electron Cyclotron Resonant Heating (ECRH), Ion Cyclotron Resonant Heating (ICRH) and Neutral Beam Heating (NBH) \cite{cit5}.
In this framework, the conceptual design of the beamline for the DTT Neutral Beam Heating system, based on negative ions, is overviewed, with a particular focus on the technical solutions adopted to fulfill the requirements and maximize beamline performances.
The proposed system features a beamline providing deuterium neutrals (D$_0$) with an energy of 510 keV and an injected power of 10 MW. Regarding the effect of the NBI in DTT, recent studies are reported in \cite{cit6} and \cite{cit7}.
An overview of the current conceptual design of the beamline for the DTT Neutral Beam Injector (NBI) is given in Fig.  \ref{Fig_general}, while Tab. \ref{Tab_functional} reports the main functional parameters.

Similarly to the NBIs of JT60 \cite{cit8} and LHD \cite{cit9}, a design with an air-insulated beam source is adopted for DTT NBI, i.e. the accelerator and ion source assemblies are connected to the rear part of the vacuum vessel. This solution was selected because it maximizes the Reliability and Availability indexes (evaluated as in \cite{cit10}), by improving the beam source accessibility and simplifying the design. The DTT NBI design differs from the Japanese scheme in the choice of the ion source: in fact it is proposed to use the same Radio Frequency source concept adopted for ITER, mainly developed by IPP Garching \cite{cit11}. The Beam Line Components (BLC), i.e. the Neutralizer, the Residual Ion Dump (RID) and the Calorimeter, will be ITER-like too, whereas the vacuum vessel will not include any large flanges (differently from ITER) to reduce cost and weight. For the vacuum pumping, it is foreseen to have a system based on turbo-molecular pumps located on the side walls of the vessel and Non-Evaporable Getter (NEG) pumps located on the upper and lower surfaces of the vessel.

More information on the conceptual design of the beamline for DTT NBI can be found in \cite{agostinetti_dtt_beamline}.

\begin{table}
\caption{Main functional parameters of the DTT NBI.}
\label{Tab_functional}
\begin{footnotesize}
\begin{center}
\begin{tabular} {|p{5.5cm}|p{1.5cm}|}
\hline
   \textbf{Parameter} & \textbf{Value} \\
\hline
Injected power & 10 MW \\
\hline
Beam Energy &	510 kV \\
\hline
Accelerated D$^-$ current	& 40 A \\
\hline
Extracted D$^-$ current density &	$>$ 239 A m$^{-2}$ \\
\hline
Ion source filling pressure	 & $\geq$ 0.3 Pa \\
\hline
Extracted current uniformity &	$\pm$10\% \\
\hline
Beam on time	& 50 s \\
\hline
Co-extracted electron fraction (e$^-$/D$^-$)	& $<$ 1 \\
\hline
Beamlet divergence    &   $<$ 7 mrad \\
\hline
Auxiliaries/extraction efficiency &	0.9 \\
\hline
Accelerator efficiency &	0.8 \\
\hline
Beam source/neutraliser entrance transmission efficiency &	0.95 \\
\hline
Neutralizer efficiency &	0.55 \\
\hline
Beam line/duct transmission efficiency	& 0.95 \\
\hline
\end{tabular}
\end{center}
\end{footnotesize}
\end{table}

        \begin{figure}
        \centerline{\includegraphics[width=0.5 \textwidth]{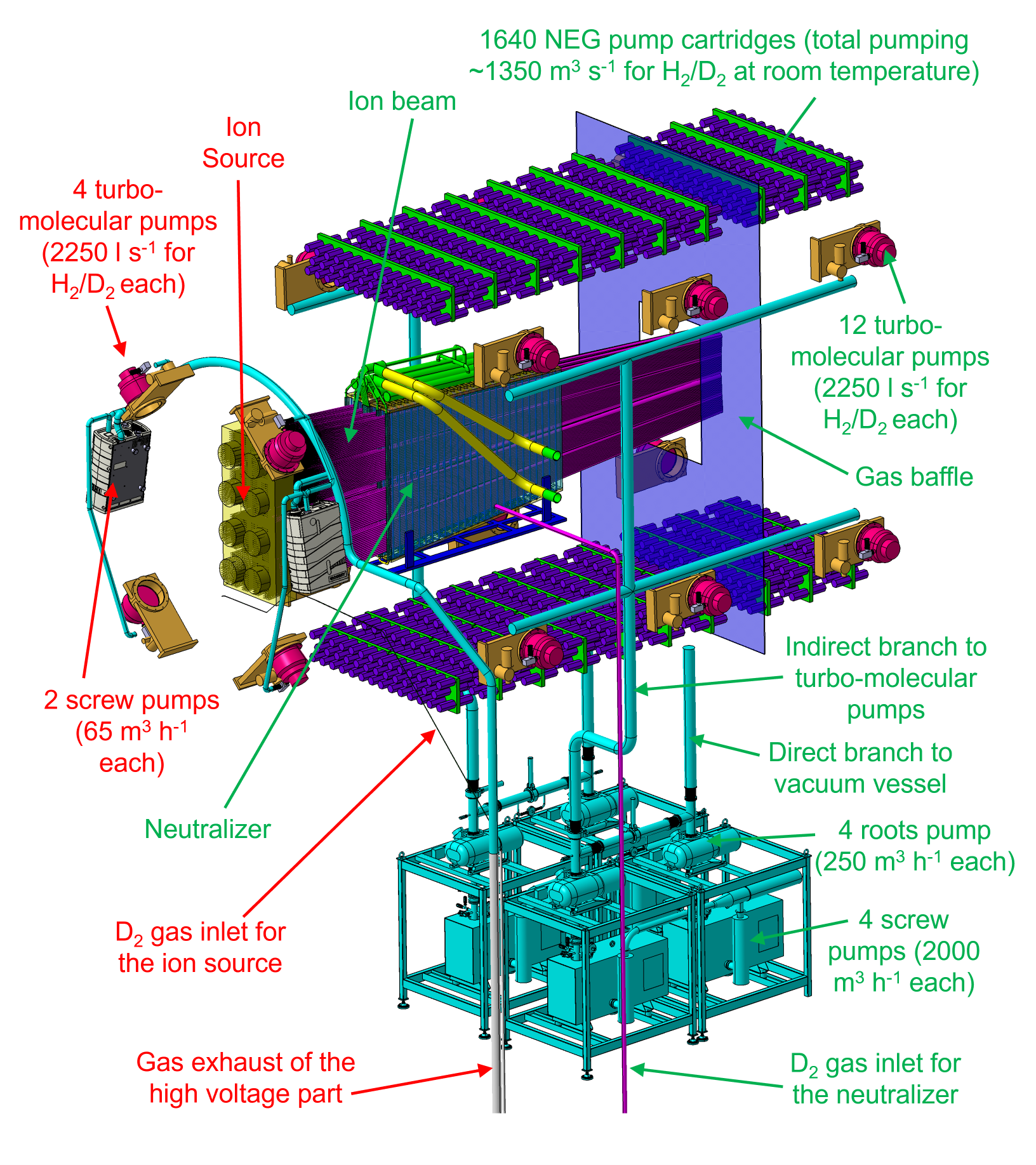}}
        \caption{Overview of the GVS for DTT NBI. Green labels: grounded section; red labels: high voltage section.}
        \label{Fig_GVS}
        \end{figure}

\section{Gas injection and Vacuum System}
\label{GVS}

The Gas injection and Vacuum System (GVS) for the DTT NBI, shown in Fig. \ref{Fig_GVS}, is composed of two systems working in parallel:
\begin{itemize}
  \item A grounded section connected to the main vacuum vessel (green labels in Fig.  \ref{Fig_GVS}). This part is made of a fore-vacuum system (given by 4 screw and 4 roots pumps) plus a high vacuum system based on 12 turbo-molecular pumps, located on the side walls of the vessel, and 1640 cartridges featuring ZAO\textsuperscript{\tiny\textregistered}\ sintered porous NEG material, located inside the vessel on the upper and lower surfaces. The function of this section is to keep the background gas density at low levels ($\sim$5$\cdot$10$^{19}$ m$^{-3}$ in the neutralizer and $\sim$10$^{18}$ m$^{-3}$ in the other regions of the vacuum vessel) and at the same time provide a suitable injection of deuterium gas from the ion source and from the neutralizer, in order to generate the ion beam and to neutralize it. To reduce the background gas density in the calorimeter and duct region (this region is particularly critical for the re-ionization losses), there is a gas baffle located between the RID and the calorimeter, as shown in Fig. \ref{Fig_GVS}.
  \item A high voltage section connected to the ion source vessel and working at -510 kV voltage (red labels in Fig.  \ref{Fig_GVS}). This part consists of a fore-vacuum system (given by two compact screw pumps mounted on the external surface of the ion source vessel) plus a high vacuum system based on 4 turbo-molecular pumps, also located on the side walls of the ion source vessel. The function of this section is to keep the background gas pressure on the back part of the ion source at very low levels (lower than 9.7 $\cdot$ 10$^{18}$ m$^{-3}$, corresponding to a pressure at room temperature of about 0.04 Pa) to avoid discharges between the RF coils and the ion source vessel, or between the RF coils.
\end{itemize}

The conceptual design of the gas injection system, fore-vacuum group and turbo-molecular pumps for DTT NBI is well consolidated as it benefits from the large R\&D and manufacturing experience of RFX \cite{dalbello}. On the other hand, the NEG technology has been applied only recently to fusion projects \cite{siviero1,siviero2,motojima,siragusa}, hence we will focus the attention on this part of the system.

        \begin{figure}
        \centerline{\includegraphics[width=0.5 \textwidth]{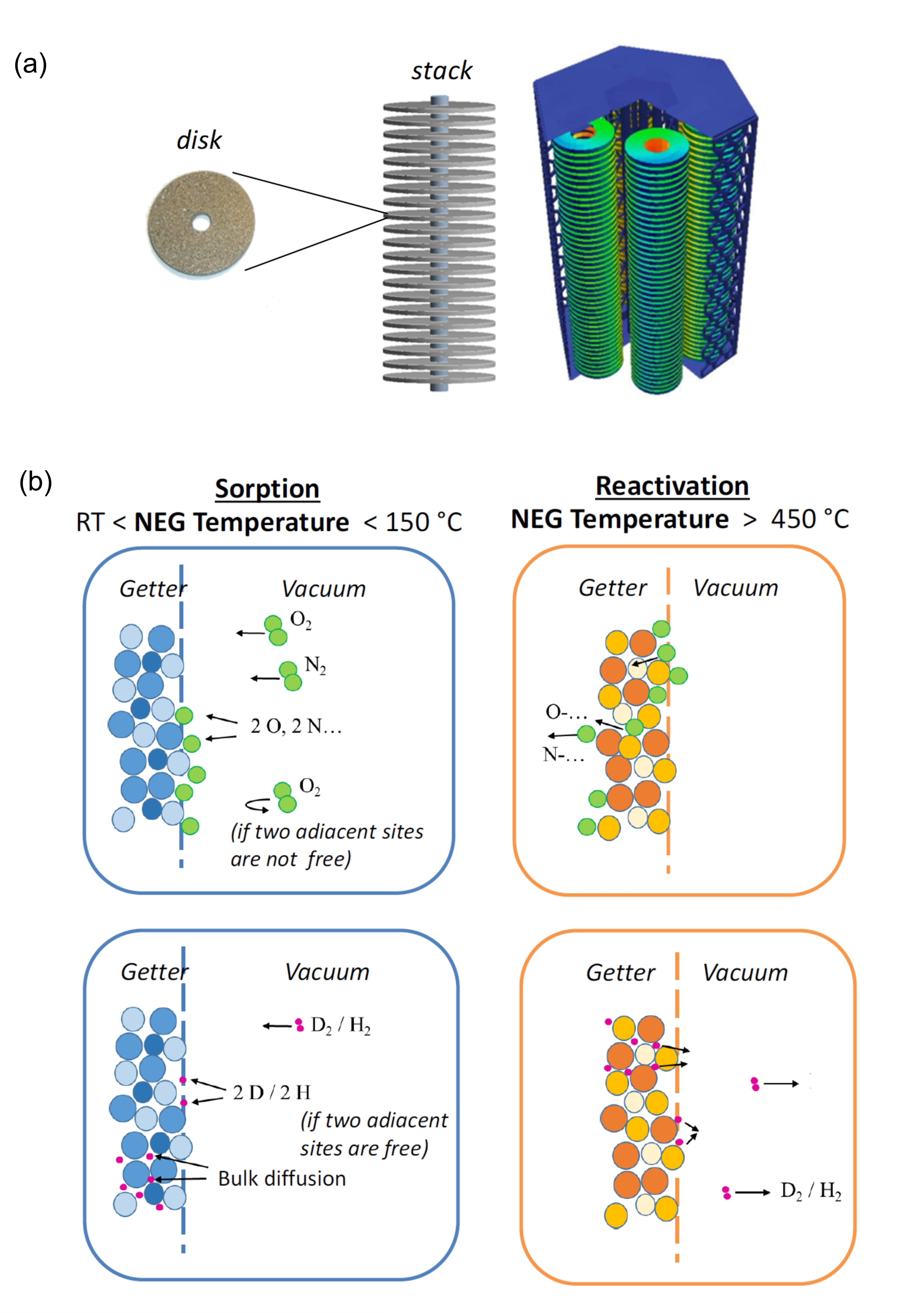}}
        \caption{NEG pumps for fusion: (a) typical design; (b) sketches of sorption and reactivation phases.}
        \label{Fig_NEG}
        \end{figure}

        \begin{figure}
        \centerline{\includegraphics[width=0.5 \textwidth]{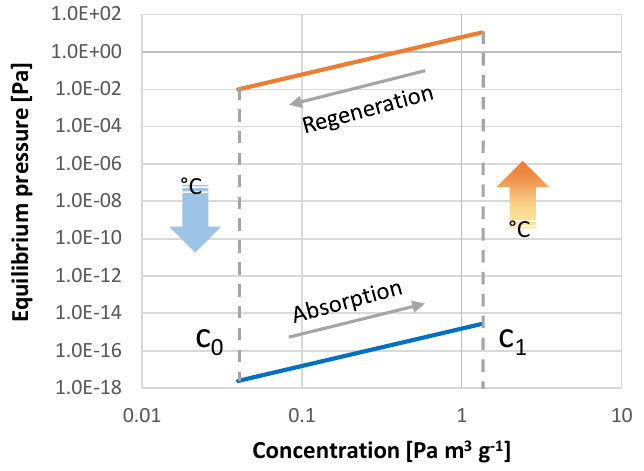}}
        \caption{Typical sorption and regeneration cycle of a NEG pump operating in hydrogen. The pressure indicated in the plot is the equilibrium pressure for hydrogen. During the absorption phase, considered at 25 $^\circ$C, the vacuum vessel pressure will be higher than the equilibrium pressure, while during the regeneration phase, considered at 550 $^\circ$C, the vacuum vessel pressure will be equal or lower than the equilibrium pressure.}
        \label{Fig_cycle}
        \end{figure}

\section{Evaluations on the application of NEG pumps in DTT NBI}
\label{evaluations}	

NEG pumps are made of ZAO\textsuperscript{\tiny\textregistered}\ getter disks stacked in a suitable configuration in order to maximize their ability to pump the surrounding environment, as shown in Fig. \ref{Fig_NEG}a.
The operation of a NEG pump foresees cycles with two subsequent phases, sketched in Fig. \ref{Fig_NEG}b:
\begin{itemize}
  \item A sorption phase (with temperature $<$ 150 °C), when the getter material absorbs the gas in the material structure.
  \item A reactivation (or regeneration) phase (with temperature $>$ 450 $^\circ$C), when the getter material releases the hydrogen in the vacuum while the other gases diffuse into the material structure.
\end{itemize}
A typical sorption-regeneration plot is shown in Fig. \ref{Fig_cycle}.

\begin{figure}
        \centerline{\includegraphics[width=0.4 \textwidth]{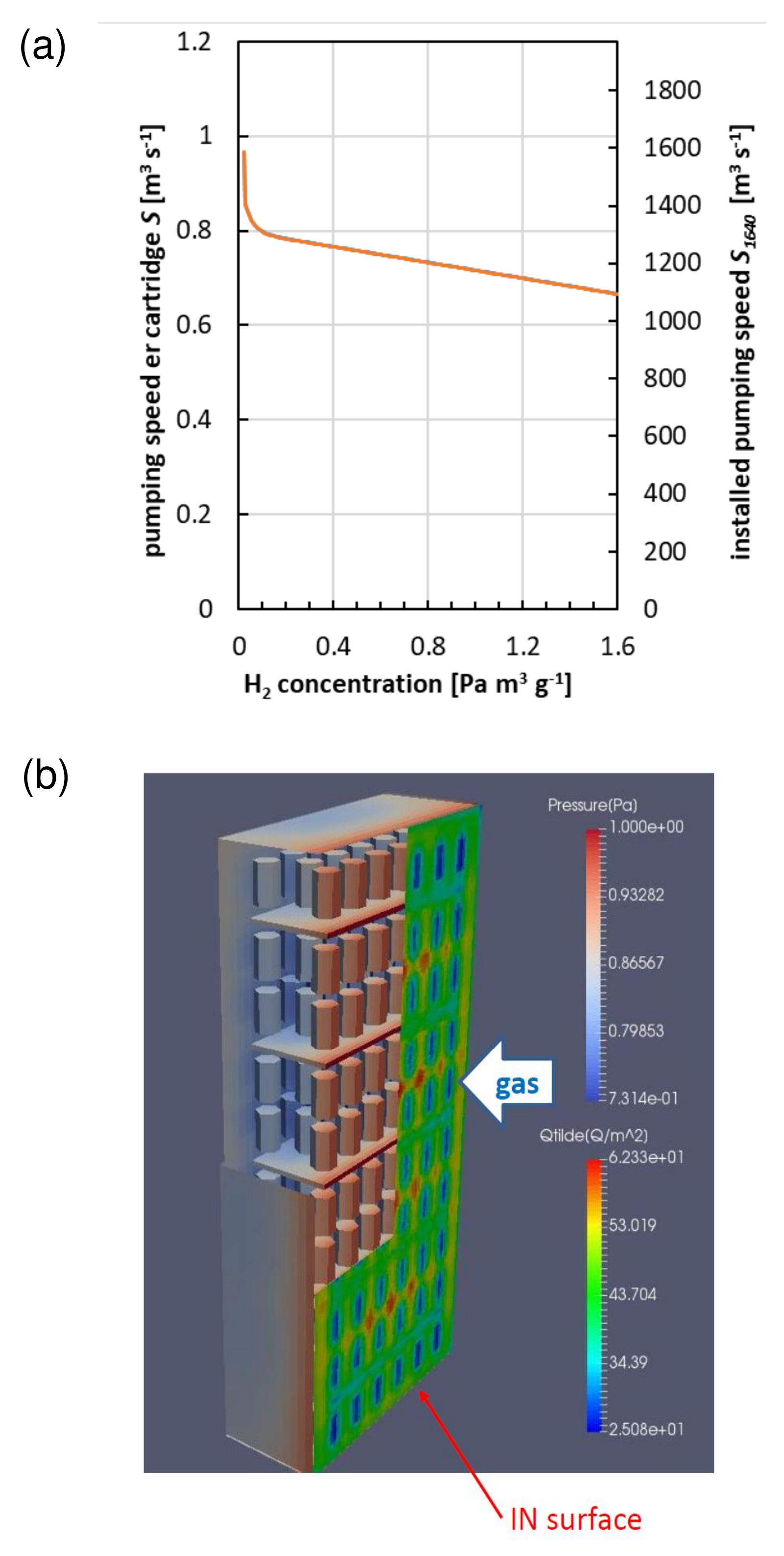}}
        \caption{Pumping speed of a ZAO NEG pump: (a) pumping speed as a function of hydrogen concentration, measured on the prototype described in \cite{siragusa}; (b) model to calculate the capture coefficient with the presently considered layout of NEG cartridges in DTT NBI. Q is the throughput measured in Pa m$^3$ s$^{-1}$, while Qtilde is the specific throughput (or throughput per unit of area) measured in Pa m s$^{-1}$. A detailed description of the model used is available in \cite{siragusa2}.}
        \label{Fig_speed}
        \end{figure}

With the most recent getter materials \cite{siviero1}, the pumping speed of the getter material is rather independent from the concentration of hydrogen in the material as can be seen from Fig.  \ref{Fig_speed}a.
With the presently considered layout of the NEG cartridges in DTT NBI, using a dedicated model with the AVOCADO code (shown in Fig.  \ref{Fig_speed}b) the capture coefficient is estimated to be around 0.1.
The capture coefficient of a gas is defined as the ratio of the number of particles captured by the pumping system to the number of total incident particles.

Other important aspects to be taken into account when designing a vacuum pumping system based on ZAO for fusion are:
\begin{itemize}
  \item The pumping speed increases linearly with operating temperature (for example, between RT and 210 °C, there is an increase of about 25\%) \cite{siviero2}.
  \item Other getterable gases (O$_2$, N$_2$, etc.) may passivate the getter surface and therefore reduce the pumping speed for H$_2$.
  \item NEG pumps can withstand a limited number of exposures to air, i.e. after about 100 exposures the pumping speed for H$_2$ could be significantly reduced.
  \item the performance of the NEG pumps decrease if water vapour at pressure $>$ 1 Pa gets in contact with the hot getter material.
  \item With hydrogen isotopes, the regeneration time is extremely dependent on the target concentration in the NEG material, i.e. it is very slow aiming at low concentrations and fast at high concentrations.
  \item The regeneration rate is also linearly correlated with the auxiliary pumping speed \cite{siviero1}.
  \item The getter material is permanently damaged if it is exposed to air during the regeneration, i.e. when the NEG pump is at high temperature (about 550 °C).
  \item No gas release is possible from the getter material, except when the getter material is heated \cite{siviero2}. On the other hand, cryopumps release all adsorbed gas if their power supply stops.
  \item Regarding the interaction with the ZAO NEG material, H$_2$ and D$_2$ behave in the same way: the sticking coefficient is very similar while the equilibrium pressure is about 20\% higher for D$_2$ \cite{siviero1}. Taking this into account, data obtained with H$_2$ can be used for the design of a device that will mostly work with D$_2$.
\end{itemize}

Analogously to the case of cryopumps, the amount of deuterium that can be stored in the NEG pumps is limited by safety reasons.
Indeed, a deuterium inventory exceeding a certain value in the NEG pumps could be dangerous, because if some air enters the vessel (for example from a broken window or a leak) there could be the risk of explosion due to the ignition of the deuterium in presence of oxygen. A fast increase of partial pressure of deuterium in the vacuum vessel could happen for example in case of a sudden release of all the deuterium sorbed by the pumps (this is valid both for systems based on cryopumps and NEG pumps).

With cryopumps, this event is quite likely. For example, an electrical power outage could make the temperature suddenly increase above cryogenic temperatures. On the other hand, with NEG pumps this event is much less likely because hydrogen isotopes obey an equilibrium pressure law (Sieverts law); at low concentrations like the ones expected for DTT NBI (not more than 0.2 Pa m$^3$ g$^{-1}$) and at temperatures as high as 900 $^\circ$C, the equilibrium pressure is still less than 50 Pa.

The most dangerous scenario for the NEG pumps is the one where the NEG pumps suddenly heat up because they get in contact with water at a pressure in the hPa range when operating at temperature $>$ 150 $^\circ$C (for example if there is an important water leak during the regeneration). In this scenario, water would  dissociate at the NEG surface: oxygen would be chemisorbed with an exothermic process, and a significant part of the hydrogen released, possibly in safety-relevant amount. This risk is avoided when operating the NEG pumps at temperature $<$ 100 $^\circ$C, hence the temperature of the pumps during the beam operations should be $<$ 100 $^\circ$C (a higher temperature would give a better pumping but would also increase the risks in case of contact with water). In any case, the risk is still present during the regeneration phase, when the NEG pumps are at high temperature (about 550-600 $^\circ$C). These considerations suggest adopting a series of countermeasures: for example, large-throughput primary pumps could be activated in case of large water leaks to pump hydrogen away, while nitrogen venting would help saturating the NEG and diluting hydrogen. Clearly, the risk of large water leaks will be minimized first.

As explained above, the sudden release of all the deuterium content of the getter is unlikely, but we may still consider putting a limit on the amount of deuterium that can be stored in the NEG pumps. The most conservative way to do this is to consider how much would be the percentage of deuterium in the vessel in case of:

\begin{itemize}
  \item event 1: a complete release of all the deuterium trapped in the NEG pumps;
  \item event 2: a contemporary complete venting of the vacuum vessel (going at atmospheric pressure).
\end{itemize}
The resulting scenario is then to be checked against the literature data on deuterium flammability conditions, which indicate that at atmospheric pressure a combustion of deuterium can occur for mixtures deuterium-air in the range between 4\% and 75\% \cite{graham}.
It should be noted that with NEG pumps both the events 1 and 2 are unlikely events, hence the probability that they happen at the same time is very low, nevertheless for these evaluations at the conceptual design stage we preferred to stay on the safe side with a large margin.

In the DTT NBI case, we are considering low values of deuterium partial pressure, hence the important limit is the 4\% one while the 75\% one is not relevant for this application. This limit can be also written in terms of maximum allowable deuterium partial pressure, i.e. 4$\cdot$10$^3$ Pa. In fact, if there is a 4$\cdot$10$^3$ Pa deuterium partial pressure inside the vessel, in case of a sudden venting to atmospheric pressure due to a leak, there will the co-existence of this 4$\cdot$10$^3$ Pa deuterium partial pressure with an air partial pressure of about 1$\cdot$10$^5$ Pa, leading to a deuterium-air mixture with a deuterium concentration of about 4\%, to be avoided because it would be at the edge of the flammability conditions.

For its experiments with cryopumps, the Lawrence Livermore Laboratories proposed an administrative limit of 1.7$\cdot$10$^3$ Pa, taking a large safety margin against the 4$\cdot$10$^3$ Pa limit \cite{graham}. To be very conservative, we will also consider this value for the following evaluations. This means assuming a margin against flammability conditions of 2.35.

The main evaluations carried out are described in the following.

        \begin{figure}
        \centerline{\includegraphics[width=0.3 \textwidth]{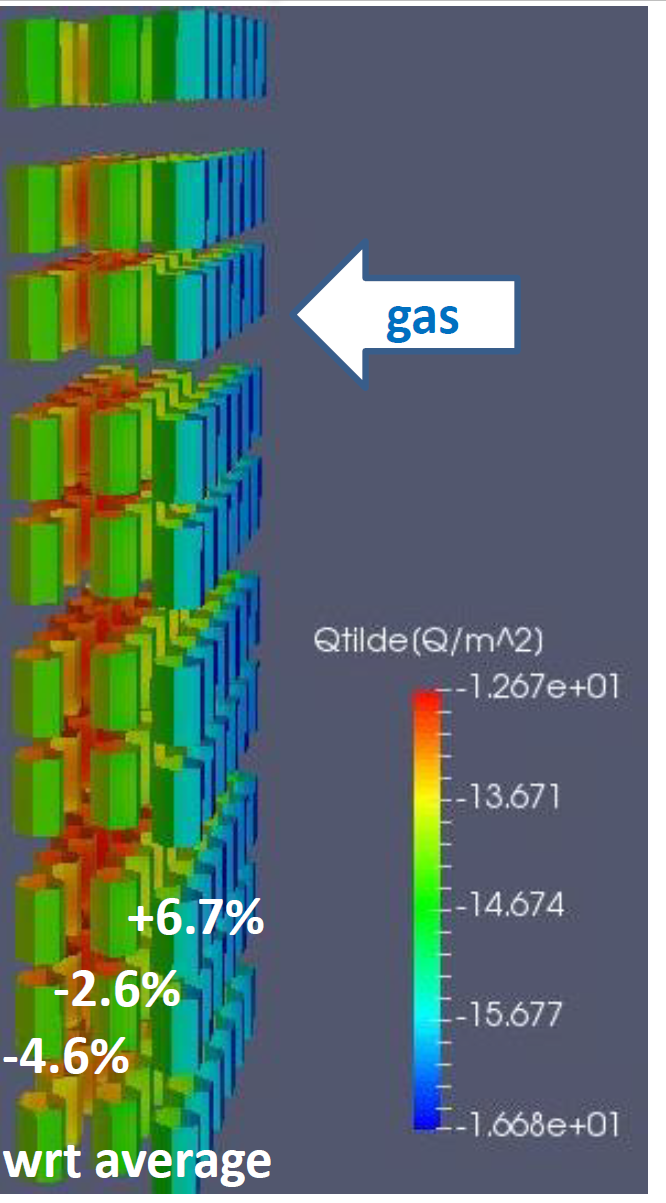}}
        \caption{Expected uniformity of gas load and effect of mutual shading between cartridges on pumping speed. Qtilde is the specific throughput (or throughput per unit of area) on the surface of the cartridges measured in Pa m s$^{-1}$, negative because it is absorbed throughput. These specific throughputs have been obtained with a pressure of 1 Pa imposed to the "In surface", as shown in Fig. \ref{Fig_speed}, but the results in term of uniformity (indicated percentage values) are valid also for other values of pressure, as the molecular regime approximation can be assumed being in vacuum conditions.}
        \label{Fig_uniformity}
        \end{figure}

\subsection{Evaluation of the maximum inventory limit}

The maximum inventory limit of deuterium in the NEG pumps can be calculated multiplying the maximum allowed deuterium partial pressure by the vessel volume. The total volume inside the vacuum vessel of DTT NBI is 6.45 m $\cdot$ 5.4 m $\cdot$ 2.96 m $\sim$ 103 m$^3$. Assuming 1.7$\cdot$10$^3$ Pa for the maximum allowed deuterium partial pressure, we obtain 1.7$\cdot$10$^3$ Pa $\cdot$ 103 m$^3$ = 175100 Pa m$^3$. This is kept as a limit for the deuterium inventory in the NEG. It is rather straightforward to stay always below this limit by controlling the amount of gas that is injected in the vessel between one regeneration and the next.

\subsection{Evaluation of the maximum deuterium concentration and check against embrittlement}
\label{concentration}
The maximum deuterium concentration in the getter material is calculated dividing the maximum inventory by the mass of getter material. As each ZAO cartridge contains 920 g of getter material, the total amount of getter material for 1640 cartridges is 1508800 g.
The layout of the cartridges has been optimized in order to obtain a high pumping capability and a good uniformity of gas concentration on the different cartridges. As shown in Fig.  \ref{Fig_uniformity}, with the currently considered layout of the NEG cartridges the distribution of absorbed D$_2$ is foreseen to be rather uniform, with a maximum difference between the cartridges of only 12\% (calculated as (106.7-95.4)/95.4 using the data of Fig. \ref{Fig_uniformity}) corresponding to a variation of $\pm$6\% from the average value. In the 1.7$\cdot$10$^3$ Pa case, the maximum hydrogen concentration is calculated as 175100 Pa m$^3$ / 1508800 g = 0.116 Pa m$^3$ g$^{-1}$. It can be observed that this concentration is 16 times lower than the embrittlement limit of the getter material, that is of 1.85 Pa m$^3$ g$^{-1}$ \cite{siviero1}. This means that the getter material will be always operating in suitable conditions and will not risk to be damaged by an excessive concentration of absorbed deuterium.

\subsection{Evaluation of the regeneration time}
The regeneration time can be then estimated considering a suitable operating time between two regenerations and using the available experimental data on regeneration. The maximum beam-on time of DTT NBI during tokamak operations is foreseen to be 50 s every hour while it is foreseen to be 50 s every 10 minutes during the conditioning phase. As a reasonable number, we considered 84 pulses between two regenerations, corresponding to about two weeks of tokamak operations (1 pulse per hour) or about two days of conditioning (6 pulses per hour). This correspond to 4200 s of operation time.
This operation time corresponds to an increase of deuterium concentration in the getter material of 0.066 Pa m$^3$ g$^{-1}$, calculated as 4200 s $\cdot$ 23.6 Pa m$^3$ s$^{-1}$ / 1508800 g, where:
\begin{itemize}
  \item 4200 s is the assumed suitable operating time between two regenerations;
  \item 23.6 Pa m$^3$ s$^{-1}$ is the total expected inlet gas throughput during beam operation, calculated as the sum of 2.8 Pa m$^3$ s$^{-1}$ injected into the ion source to generate the negative ions (D$^-$) plus 20.8 Pa m$^3$ s$^{-1}$ injected into the neutralizer to transform the negative ions (D$^-$) into neutrals (D$^0$);
  \item 1508800 g is the total amount of getter material for 1640 cartridges;
\end{itemize}

\begin{figure}
        \centerline{\includegraphics[width=0.5 \textwidth]{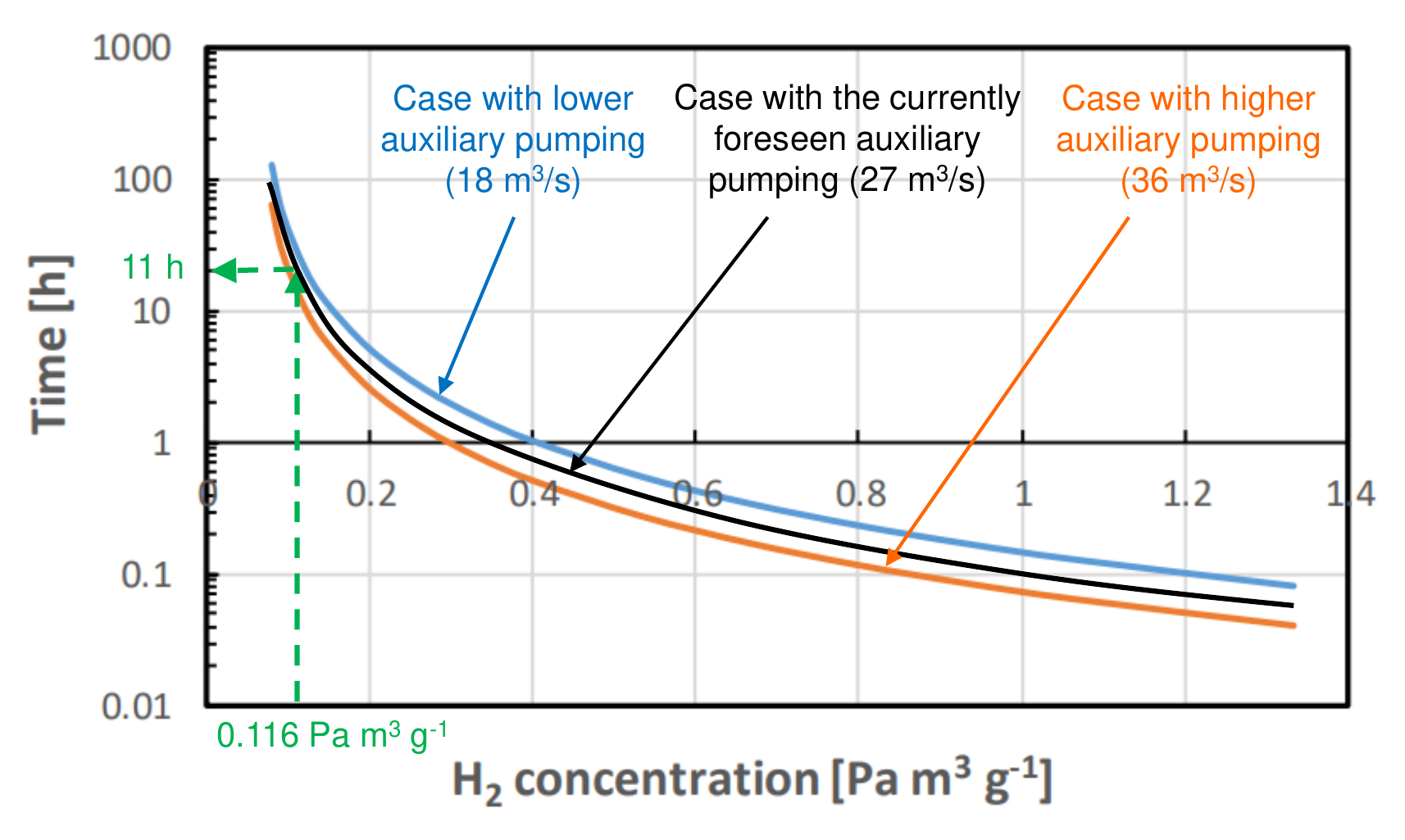}}
        \caption{Foreseen deuterium regeneration time of the getter material in DTT NBI as a function of hydrogen concentration, based on the experimental data on the prototype of NEG application for fusion \cite{siragusa}. Regeneration temperature: 600 °C. The auxiliary pumping has been assumed of 27 m$^3$ s$^-1$ as in the current design of DTT NBI, nevertheless also the curves at 18 and 36 m$^3$ s$^-1$ are shown for a sensitivity analysis.}
        \label{Fig_time}
        \end{figure}

Hence, 0.066  Pa m$^3$ g$^{-1}$ is the typical amount of deuterium concentration to be removed from the getter material during every regeneration cycle. In the considered 1.7$\cdot$10$^3$ Pa case, removing 0.066 Pa m$^3$ g$^{-1}$ (corresponding to 4200 s of operation time) means decreasing the deuterium concentration in the getter material from 0.116  Pa m$^3$ g$^{-1}$ (the maximum allowed concentration based on Par. \ref{concentration}) to 0.05 Pa m$^3$ g$^{-1}$. Using experimental data on regeneration (as shown in Fig.  \ref{Fig_time}), it can be estimated that this process needs about 11 hours at 600 °C.  Adding 1 hour for heating and 4 hours for cooling the NEG pumps, in the 1.7$\cdot$10$^3$ Pa case the total time for a whole regeneration process (upper path between C$_1$ and C$_0$ in Fig. \ref{Fig_cycle}) is estimated to be about 16 hours. This time for the whole regeneration process is considered to be reasonable.

\section{Influence on transmitted power}

One of the most significant parameters to be checked during the development of the DTT NBI design is the neutral beam power that is able to reach the plasma in the tokamak. Indeed, a fraction of the power is lost due to beam interception and another fraction to beam reactions. The latter ones are strongly depending on the density of the background gas in the vacuum vessel, which in turn depends on the GVS.

The GVS has a strong influence on this parameter. In fact, if the vacuum pumping is not sufficiently good in keeping the density of background gas low in the NBI vessel, the amount of power reaching the plasma decreases significantly because the power losses due to stripping reactions of the charged beam in the accelerator and re-ionization reactions of the neutral beam in the calorimeter and duct regions increase.

This means that the vacuum pumping system is requested to maintain the density of the background gas inside the vessel at very low values ($\sim$5$\cdot$10$^{19}$ m$^{-3}$ in the neutralizer and $\sim$10$^{18}$ m$^{-3}$ in the other regions of the vacuum vessel) in presence of a direct injection of deuterium gas from the ion source and from the neutralizer, needed to generate the ion beam and to neutralize it.

        \begin{figure}
        \centerline{\includegraphics[width=0.5 \textwidth]{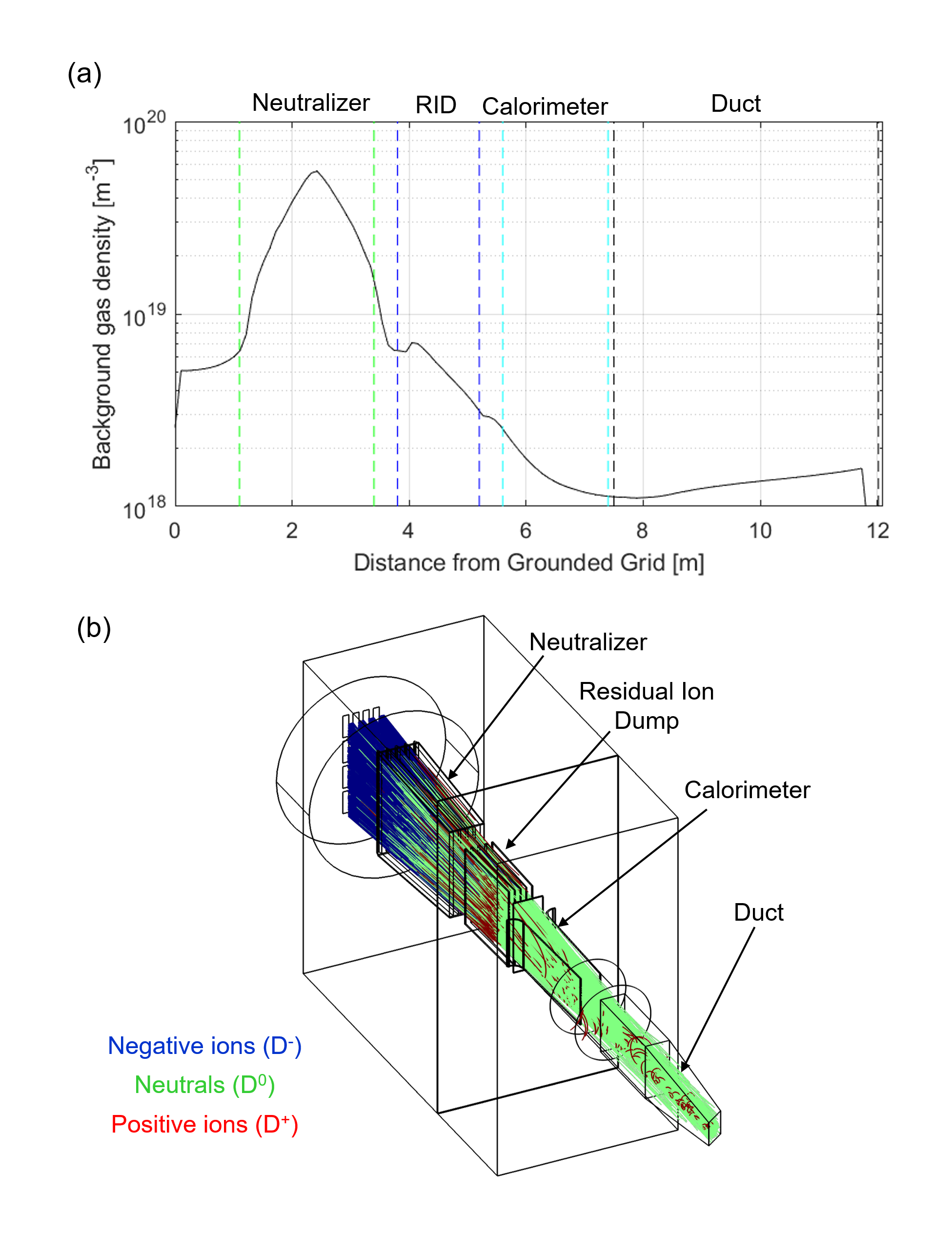}}
        \caption{Vacuum evaluations and effect on the beam: (a) typical profile of the background gas density along the beam direction; (b) typical trajectories of the particles.}
        \label{Fig_plots}
        \end{figure}

For this reason, the performances of the vacuum system must be extremely good in terms of pumping capability.
Hence, all the proposed design options were evaluated using a suite of multi-physics models developed using the commercial FEA code COMSOL, merging vacuum pumping evaluations, electromagnetic phenomena and beam reactions with background gas to determine the heat loads along the beamline and the power transmitted to the plasma.

The design solution here presented for the GVS allows to obtain an amount of power reaching the plasma in the tokamak in line with the expectations, as shown in Tab. \ref{Tab_power}. In fact, the power reaching the plasma is higher than the required value of 10 MW.

The corresponding profile of the background gas density, together with a typical plot of the particle trajectories inside the beamline, is shown in Fig. \ref{Fig_plots}. This simulation was carried out considering deuterium gas and simulating the NEG pumps by assuming a 0.1 capture coefficient on the upper and lower surface of the vessel.

\begin{table}
\caption{Intercepted and transmitted power with the current GVS design.}
\label{Tab_power}
\begin{footnotesize}
\begin{center}
\begin{tabular} {|p{3.5cm}|p{1.5cm}|}
\hline
   \textbf{Region} & \textbf{Power [MW]} \\
\hline
Neutralizer & 2.1 \\
\hline
Residual Ion Dump &	9.2 \\
\hline
Calorimeter (open)	& 0.5 \\
\hline
Duct &	0.3 \\
\hline
Tokamak plasma	 & 10.5 \\
\hline
\end{tabular}
\end{center}
\end{footnotesize}
\end{table}

\section*{Conclusions}

A conceptual design for a Gas injection and Vacuum System (GVS) has been developed for DTT NBI.
The current solution features two parts, one at ground potential and one at high voltage. The part at ground potential consists of a conventional fore-vacuum system, made of screw and roots pumps, plus a high vacuum section made of turbo-molecular pumps and ZAO based getter pumps. The part at high voltage features compact screw pumps and turbo-molecular pumps.
The design has been optimized in terms of the most relevant aspects, i.e. the pumping capability, the margin against danger of deuterium ignition and the margin against embrittlement.
Based on the preliminary study carried out so far, it appears that the usage of the NEG pumps in DTT NBI could be possible, as all the DTT NBI requirements could be achieved with this system.
More detailed studies on the optimization of cartridge number and layout, thermal aspects, maintenance strategy, electrical connections and mechanical supports will be carried out in the next future.

\section*{Acknowledgments}

This work has been carried out within the framework of the EUROfusion Consortium, funded by the European Union via the Euratom Research and Training Programme (Grant Agreement No 101052200 - EUROfusion). Views and opinions expressed are however those of the author(s) only and do not necessarily reflect those of the European Union or the European Commission. Neither the European Union nor the European Commission can be held responsible for them.

\end{document}